\documentclass[lineno]{jfm}

\usepackage{graphicx}
\usepackage{newtxtext}
\usepackage{newtxmath}
\usepackage{natbib}
\usepackage{hyperref}
\usepackage{layouts}
\usepackage{caption}
\usepackage{subcaption}
\usepackage{float}
\usepackage{comment}
\usepackage{xcolor}
\usepackage{color}
\usepackage{tikz}
\usetikzlibrary{shapes}
\usepackage{gensymb}
\usepackage{graphicx,caption}
\hypersetup{
    colorlinks = true,
    urlcolor   = blue,
    citecolor  = black,
}

\newcommand{\RomanNumeralCaps}[1]
\linenumbers

\title{A family of adverse pressure gradient turbulent boundary layers with upstream favorable pressure gradients}

\author{Aadhy Parthasarathy\aff{1}
  \corresp{\email{aadhysp2@illinois.edu}},
  \and Theresa Saxton-Fox\aff{1}}

\affiliation{\aff{1}Department of Aerospace Engineering, University of Illinois at Urbana-Champaign}

\begin{document}
\newcommand{\markerone}{\raisebox{0.5pt}{\tikz{\node[draw,scale=0.3,circle,fill=black!20!black](){};}}}

\newcommand{\markergray}{\raisebox{0.8pt}{\tikz{\node[draw,gray,scale=0.2,circle,fill=gray!40!gray](){};}}}
\newcommand{\markerblack}{\raisebox{0.8pt}{\tikz{\node[draw,scale=0.2,circle,fill=black!20!black](){};}}}
\newcommand{\markerred}{\raisebox{0.8pt}{\tikz{\node[draw,red,scale=0.2,circle,fill=red!20!red](){};}}}
\maketitle

\begin{abstract}

A flat plate turbulent boundary layer (TBL) is experimentally subjected to a family of 22 favorable-adverse pressure gradients (FAPGs) by using a ceiling panel of variable convex curvature. We define a FAPG as a sequence of streamwise pressure gradients in the order of favorable followed by adverse, similar to the pressure gradient sequence over the suction side of an airfoil. The adverse pressure gradient region of this configuration is studied using particle image velocimetry (PIV) in the streamwise--wall-normal plane. The streamwise mean and Reynolds stresses are presented and discussed in this work. Although favorable and adverse pressure gradients are known to evoke opposing responses from a TBL, the current adverse pressure gradient (APG) that followed a favorable pressure gradient (FPG) was not found to `reverse' the effect of the FPG. Instead, a complex response that is neither characteristic of a ZPG nor an APG TBL was observed. The Reynolds stresses showed a bimodal structure, with the outer-scaled stresses strengthening along the streamwise direction in $y < ~ 0.2 \delta$ and weakening in $y > ~ 0.2\delta$, where $y$ is the wall-normal coordinate, and $\delta$, the local boundary layer thickness. These were signatures of a new internal layer, previously observed in bump/hill flows, formed due to the pressure gradient sign change from favorable to adverse. The non-typical structure of and trends in the statistics were more significant in the stronger pressure gradient cases, where the stronger upstream FPG seemed to have exerted a stronger influence on the downstream APG. The current dataset with 22 cases provided a well-resolved picture of the breakdown of equilibrium of the ZPG TBL under an FAPG configuration. The statistics for all the cases are included under the ancillary file section.

\end{abstract}


\section{Introduction}
\label{sec:intro}

Pressure gradients add significant complexity to the physics of turbulent boundary layers (TBLs). Most of the complexity stems from the strong dependence of the flow on upstream conditions, dubbed `history effects' \citep{vinuesa2017revisiting}. Rather than the local value of non-dimensional pressure gradient, the cumulative effect of its upstream variation governs the statistics and structure of the TBL at that streamwise station considered. This makes it difficult to ascertain the generality of observations when the pressure gradient history in an experiment or simulation is not specifically-controlled. A controlled, well-defined history results in a canonical construction of the pressure gradient TBL where the Clauser's pressure gradient parameter, $\beta$, is held constant. The resulting TBL is said to be in near-equilibrium, allowing the formulation of simple scaling laws and empirical generalizations \citep{vila2020experimental,pozuelo2022adverse}. While such simplifications are undoubtedly useful, it is also worth studying the TBL under more complex pressure gradients that engineering situations often present with, to accelerate the accurate modelling and prediction of these flows. 

A particularly interesting class of complex pressure gradient TBLs result when the imposed pressure gradients alternate in sign between favorable (negative) and adverse (positive). Favorable and adverse pressure gradients evoke essentially the opposite responses from a TBL: favorable accelerates the boundary layer, creating enlarged viscous and buffer layers and a thinned-down wake region, causes a significant suppression of turbulent stresses, especially in the outer region, and pushes the boundary layer to become laminarescent for a strong enough FPG; adverse decelerates the boundary layer, diminishes the viscous/buffer layers and  enlarges the wake region, significantly energizes the outer region turbulence, and pushes the boundary layer to separate for a strong enough APG. When these conditions act sequentially, it is less clear if and how their effects cumulate. The limited studies that have probed such configurations report deviations from well-known pressure gradient effects of the mean, turbulent stresses, and the skin friction. These include observations of a more rapid departure from standard law behaviors compared to when a continuously adverse or favorable pressure gradient of similar strength is imposed, and the appearance of multiple knee points in the Reynolds stress profiles \citep{webster1996turbulence,cavar2011investigation}. Asymmetric recovery depending on the exact sequence of favorable-adverse pressure gradients have also been observed \citep{bandyopadhyay1993turbulent}. The flow over a bump / hill is an example of such a complex pressure gradient imposition, imposing a sequence of mild APG at the foot, strong FPG until the apex, strong APG in the downstream half, and a mild FPG at the end of the bump / hill \citep{balin2021direct,uzun2021simulation}. Globally, the stabilizing effect of the strong FPG until the bump apex has been shown to cause the APG TBL to become more resilient to flow separation \citep{balin2021direct,webster1996turbulence}. These studies have unanimously highlighted the inadequacy of lower-fidelity simulations in sufficiently accurately predicting several aspects of the TBL response, due especially to the inability of turbulence models to replicate FPG effects and correctly carry it forward into the succeeding APG region. These shortcomings are well-summarized in \citet{matai2018flow} and some of the on-going validation efforts involving experiments and high-fidelity simulations of TBLs encountering pressure gradient sequences are presented in \cite{slotnick2019integrated}.

The present experimental investigation is aimed at gaining fundamental insights into pressure gradient TBLs with complex spatial history. Specifically, a favorable-adverse pressure gradient (FAPG) sequence is imposed on a flat plate TBL and the pressure gradient strength is statically increased from ZPG to a strong FAPG through 22 cases. This results in a series or ``family'' of FAPG TBLs. The APG region of the pressure gradient sequence is the focus of this work. The relatively recent studies in this area, some of which were referenced above, have been more numerical in nature than experimental, and the systematically-acquired data in the present work attempts to redress this disproportion.  

\section{Experimental Framework}\label{sec:expt}
\subsection{Unsteady Pressure Gradients Facility} \label{sec:upgf}

\begin{figure}
    \centering
    \captionsetup{width=1\linewidth}
    \includegraphics[scale = 1,trim={8.1cm 0 0 0},clip]{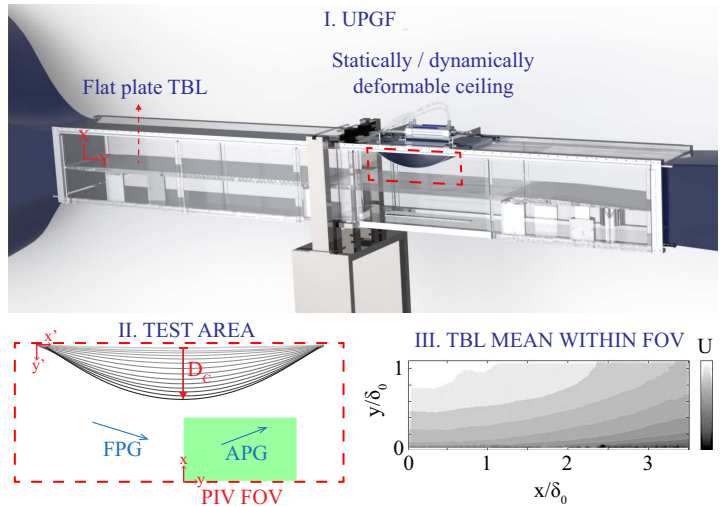}
    \caption{Illustration of the experimental details. I. Unsteady pressure gradients facility at UIUC. The red box bounds the test area. II. Close-up view of the test area illustrating the family of favorable-adverse pressure gradients being imposed on the flat plate. $D_c$ is the vertical distance between the flat ceiling and any deflected ceiling. The field of view for PIV is set in the APG region and is marked by the green box. III: Example TBL mean computed from PIV}
    \label{fig:expt}
\end{figure}

Experiments were conducted in the Unsteady Pressure Gradients Facility (UPGF), located at the UIUC Aerodynamics Research Laboratory (figure \ref{fig:expt}). The UPGF comprises of a boundary layer wind tunnel and a removable installation to generate steady and unsteady streamwise pressure gradients \citep{parthasarathy2022novel}. The wind tunnel accesses low subsonic speeds in the range 1-40 m/s within a test section of dimensions $0.381 \times 0.381 \times 3.66$ m. Inlet flow conditioning is achieved through honeycomb straighteners, turbulence-reducing screens, and a high contraction ratio (27:1). A flat plate with a leading edge trip is mounted within the test section spanning its entire length in order to develop a nominally-ZPG TBL. The freestream conditions relevant to the current tests are summarized in table \ref{tab:freestream}, all measured at the center of the test area, including the freestream velocity, $U_0$, $99\%$ boundary layer thickness, $\delta_0$, displacement thickness, $\delta_0^*$, freestream turbulence intensity, $TI$, friction velocity, $u_{\tau_0}$, Reynolds number based on distance from the leading edge, $\Rey_{X_0}$, and friction Reynolds number, $Re_{\tau_0}$. $u_{\tau_0}$ was computed by applying the Clauser method \citep{clauser1956turbulent} on the the ZPG mean velocity profile obtained from PIV. 

At a distance of 2.35 m from the leading edge of the flat plate, a 0.61 m section of the ceiling is replaced with the pressure gradients installation. The installation holds a flexible metal ceiling panel within the test section that is connected mechanically to an actuation mechanism above the test section. The mechanism is used to deform the ceiling panel to the shape of an inverted convex bump of different radii, imposing favorable-adverse pressure gradient sequences of different strengths. The ceiling deformation can be performed either statically or dynamically, depending on whether steady or unsteady pressure gradients are desired. The details of this setup can be found in \citep{parthasarathy2022novel}. For the present work, a series of steady pressure gradients were generated by statically deforming the ceiling to 22 different curvatures. The vertical extent of deformation ($D_c$ in figure \ref{fig:expt}) governs the spatial strength of the pressure gradient imposed and was set to span [1 - 78 mm] in the current 22 cases. The maximum deflection of $D_c$ = 78 mm corresponds to a minimum area ratio $(A/A_0)_{min}$ of 40$\%$, where $A$ is the cross-sectional area local to a streamwise location and $A_0$ is that upstream of the test region.

\begin{figure}
    \centering
    \captionsetup{width=1\linewidth}
    \includegraphics{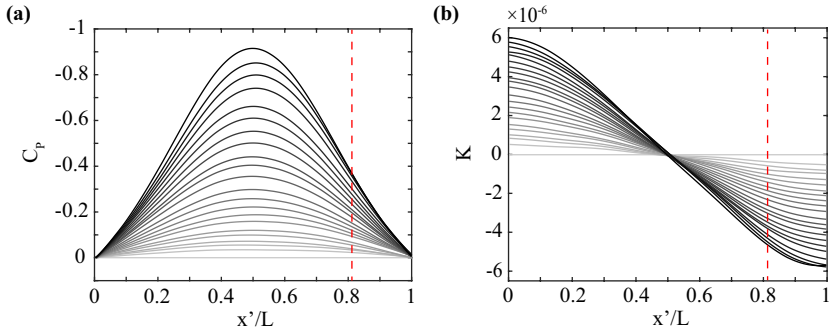}
    \caption{(a) Coefficient of pressure distributions caused by the 22 deflected ceiling states. Darker grays correspond to higher $D_c$. (b) Corresponding pressure gradient distributions, shown in terms of the acceleration parameter, $K$. The red dashed line indicates the location of flow separation from the ceiling.}
    \label{fig:pressure}
\end{figure}

The spatial distributions of the pressure coefficient, $C_P$, created in the test area due to the 22 deformed states of the ceiling are shown in figure \ref{fig:pressure}a. The profiles were computed analytically with a steady, incompressible, 1D flow assumption, using the exact geometric states of the ceiling obtained by imaging the deformed states, and were experimentally validated to be accurate within $6\%$ using high-frequency pressure measurements, the details of which can be found in \citet{parthasarathy2022novel}. In $0<x'/L<0.5$, the pressure gradient is favorable, and in $0.5<x'/L\leq0.81$, the pressure gradient is adverse. The flow over the ceiling was found to have separated in $x'/L>0.81$, marked by the red dashed line, rendering the pressure distributions invalid after this point. The FOV for PIV was chosen to remain within $x'/L \leq 0.81$. The pressure gradient variations in the test area are shown in figure \ref{fig:pressure}b in terms of the acceleration parameter, $K$ ($\equiv \frac{\nu}{U_l^2}\frac{dU_l}{dx}$, where $U_l$ is the local average velocity outside the boundary layer), valid in $0<x'/L\leq0.81$. 

\begin{table}
  \begin{center}
\def~{\hphantom{0}}
  \begin{tabular}{lccccccc}
      $U_0$ (m/s)  &   $u_{\tau_0}$ (m/s)  & $\Rey_{X_0}$   &   $Re_{\tau_0}$ & $\delta_0$ (m) & $\delta^*_0$ (m) & $TI$ (\%) \\[3pt]
         \hspace{3mm}  7.6   & 0.34 & 1.23 $\times 10^6$ & 990 & 0.042 & 0.0085 & 0.5 \\
  \end{tabular}
  \caption{Freestream conditions measured at the center of the test area}
  \label{tab:freestream}
  \end{center}
\end{table}

\subsection{Particle Image Velocimetry}\label{sec:piv}

\begin{figure}
    \centering
    \includegraphics{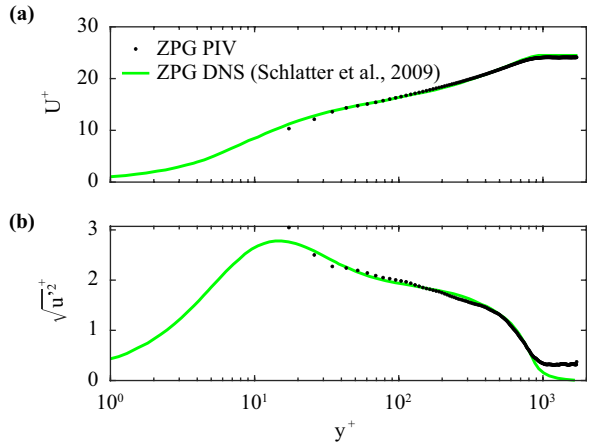}
        \captionsetup{width=1\linewidth}
    \caption{Comparison of experimental data with benchmark DNS data \citep{schlatter2009turbulent} (a) streamwise velocity (b) streamwise RMS velocity. The first three data points have higher uncertainties.}
    \label{fig:dnscomp}
\end{figure}

The response of the flat plate TBL to the pressure gradients imposed by the UPGF were captured using planar particle image velocimetry (2D-PIV) in a $150 \times 93.75$ mm ($L_x \times L_y$, $3.57\delta_0 \times 2.23\delta_0$) streamwise -- wall-normal plane set in the APG region starting at $x' = 0.228$ or $x'/L = 0.5$. This field of view (FOV) is indicated by the green box in figure \ref{fig:expt}. A mineral-oil based seeding was introduced at the tunnel inlet and a Terra PIV 527-80-M dual-pulse laser was used along with a set of sheet-forming optics to illuminate the FOV. A Phantom VEO 710L camera was used to capture the particle image pairs in a frame straddling mode. For each of the 22 steady pressure gradient impositions, 10 000 image pairs were acquired at a rate of $0.2$ kHz. The vector fields were processed using DaVis $10.1$ software using a multi-pass approach with a final interrogation window size of $16 \times 16$. The resulting vector fields had a spatial resolution of $\Delta l^+$ = 8.9. The kinematic viscosity, $\nu$, and the friction velocity, $u_{\tau_0}$ were used in defining the viscous scales. A comparison of the measured ZPG mean and streamwise RMS velocity to DNS of \cite{schlatter2009turbulent} is shown in figure \ref{fig:dnscomp}. 

\section{Results and discussion}\label{sec:results}

\begin{figure}
    \centering
    \captionsetup{width=1\linewidth}
    \includegraphics{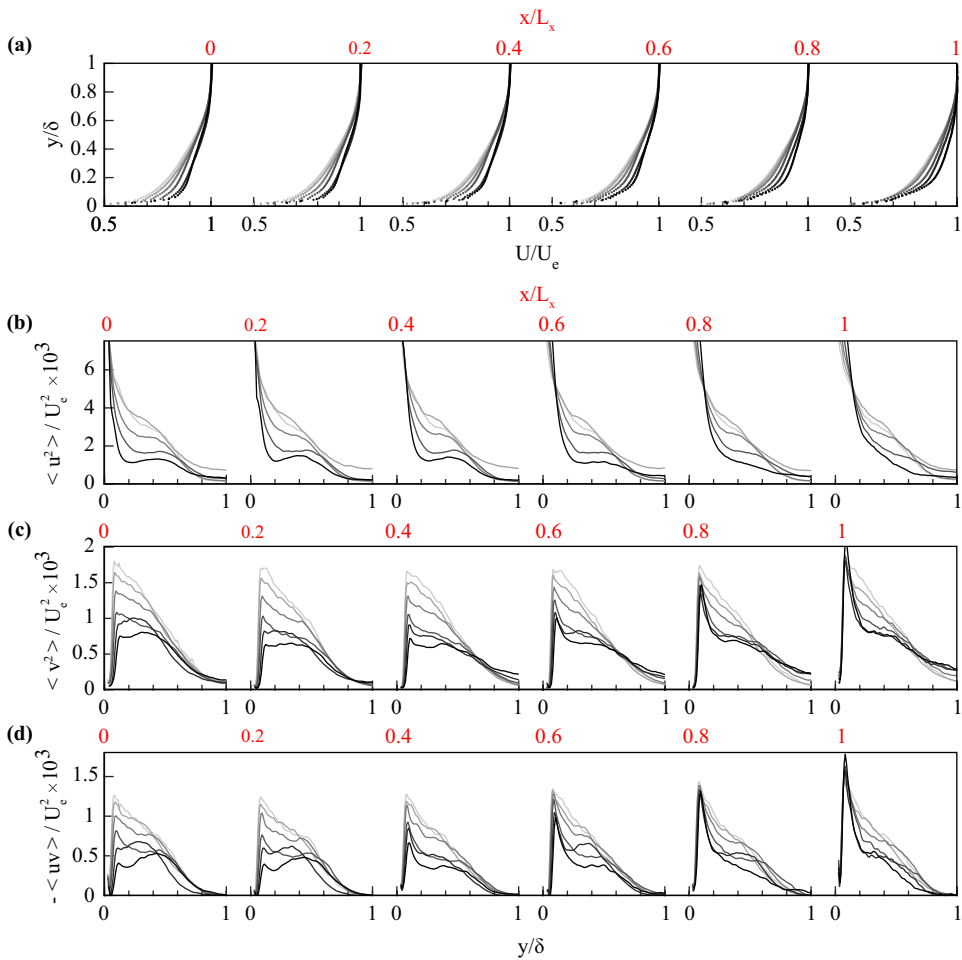}
    \caption{Statistics as a function of space ($x$) and pressure gradient ($K^*$). (a) Mean streamwise velocity (b) Streamwise Reynolds stress (c) Wall-normal Reynolds stress (d) Reynolds shear stress. The streamwise location from which the statistics are extracted is indicated above each panel in red font. Stronger pressure gradients or higher $K^*$ are marked using darker grays. $K^*$ = 0, 0.22, 0.41, 0.64, 0.8, and 1 for the cases shown.}
    \label{fig:steady_stats}
\end{figure}

The mean and Reynolds stresses of the TBL under the imposed pressure gradients are presented and discussed for 6 cases out of the 22 sets. Statistics for all the cases are made available through the supplementary material associated with this paper. The 6 cases presented here correspond to sets [1, 5, 9, 13, 17, 22], having $K$ at $x' = 0$ = [0, 1.31, 2.43, 3.75, 4.77, 5.97] $\times 10^{-6}$. Since the pressure gradient also increases in strength within the FOV along $x$, the increase with change in $D_c$ (i.e, from case 1 - 22) is referenced using an arbitrary variable, $K^*$ $\equiv$ $K/K_{max}$ = [0, 0.22, 0.41, 0.63, 0.8, 1]. 

The streamwise mean velocity, streamwise Reynolds stress (u-RS), wall-normal Reynolds stress (v-RS), and Reynolds shear stress (uv-RS) are shown in figure \ref{fig:steady_stats}, scaled with local outer-units obtained using the diagnostic plot technique for pressure gradient TBLs \citep{vinuesa2016determining}. Uncertainties in the statistics were computed with 95$\%$ confidence, following \citet{benedict1996towards},
at 2 spatial locations ($x/L_x$ = 0 and $x/L_x$ = 1) and averaged. The uncertainties were then averaged along the wall-normal direction and are as follows, in the format `Average [minimum, maximum]': in the streamwise mean, 0.21$\%$ [0.08$\%$,0.82$\%$]; in the u-RS, 3.06$\%$ [2.22$\%$,3.63$\%$]; in the v-RS, 3.38$\%$ [2.86$\%$,4.14$\%$]; in the uv-RS, 3.46$\%$ [2.47$\%$,10.18$\%$]. The streamwise locations from which the data shown in figure \ref{fig:steady_stats} are extracted are indicated above each panel in terms of $x/L_x$. $x = 0$ at the upstream end of the PIV FOV and $x = L_x$ at the downstream end. 

The mean velocity profiles show evidence of upstream acceleration from the FPG region and local deceleration from that accelerated state throughout the APG region. At $x/L_x$ = 0, at the transition from FPG to APG, the mean velocity profiles shown in figure \ref{fig:steady_stats}a are seen to become fuller for increasing pressure gradient strength ($K^*$, darker data points), an effect of acceleration caused by the upstream FPG being stronger for higher $K^*$. The mean velocities at this station resemble profiles extracted from successive spatial stations in a region of FPG \citep{ichimiya1998properties}. At all locations except $x/L_x$ = 0, the shown data are in a local APG. Moving along $x/L_x$ at a given $K^*$, the mean velocity gradient in $y < 0.2\delta$ decreases due to the deceleration caused by the APG. For a given location within $x/L_x >$ 0, the mean profiles become fuller with increasing $K^*$, despite higher $K^*$ corresponding to a stronger APG in $x/L_x >$ 0. This is because for higher $K^*$, the upstream FPG in the region $x/L_x <$  0 is also stronger, creating a more accelerated incoming TBL that the APG locally acts on. Comparing $x/L_x = 0$ and $x/L_x = 1$, the mean velocity profiles are observed to decelerate through the APG region, but to remain more accelerated than the equivalent ZPG case. This is indeed expected from the $C_P$ distributions of figure \ref{fig:pressure}a, where $C_P$ remains negative through the APG region, much like the suction side of typical airfoils at moderate angles of attack, where the strong upstream FPG causes the APG flow to remain faster than the incoming flow despite the local deceleration. Mean velocity variations under other favorable-adverse pressure gradient sequences also exhibit similar trends in the mean, such as in the flow over a bump \citep{cavar2011investigation}.

The Reynolds stresses (panels b-d of figure \ref{fig:steady_stats}) show two main trends. First, suppression and the appearance of a bimodal structure coming out of the spatially-varying FPG, and second, the strengthening of the near wall peak across the APG region. Both behaviors show traits specific to the spatially-varying pressure gradient used in this experiment. The u-RS at the transition from FPG to APG, $x/L_x$ = 0, in figure \ref{fig:steady_stats}b, exhibits a suppression of the stress throughout the measured boundary layer with increasing $K^*$, as is expected of an accelerated TBL \citep{volino2020non}. However, a two-peak structure appears, with the two peaks separated by a `knee point' at their valley. At $x/L_x=0$, the knee point can be quantified as a local minimum for the three largest pressure gradients shown ($K^* \geq$ 0.64). For these pressure gradients, the wall-normal location of the knee point reduces with increasing pressure gradient, forming at $y = 0.36 \delta$ when $K^*$ = 0.64 and at $y = 0.23 \delta$ when $K^*$ = 1. At the same streamwise location, $x/L_x = 0$, v-RS and uv-RS shown in figures \ref{fig:steady_stats}c and \ref{fig:steady_stats}d, respectively, also show a suppression of stresses caused by the upstream FPG for increasing $K^*$ and the formation of a two-peak structure. For large pressure gradients, $K^* \geq$ 0.8, the magnitude of the second peak exceeds that of the first peak. The wall-normal location of the local minimum approximately coincides with that formed in the u-RS (figure \ref{fig:steady_stats}b). The wall-normal location of the first peak in v-RS and uv-RS  at $x/L_x$ = 0 shifts away from the wall as $K^*$ increases, from $y = 0.08 \delta$ at $K^*=0$ to $y = 0.12 \delta$ at $K^*=1$.

Along the streamwise direction, for a given $K^*$, the knee point generally persists and its wall-normal location moves away from the wall. The knee point in the uv-RS is located at $y = 0.28 \delta$ for $K^*$ = 1 and $x/L_x=1$, versus $y=0.23\delta$ at $x/L_x=0$.  In $x/L_x >$  0, the peak moves closer to the wall and reverts back to $y = 0.08 \delta$ at $x/L_x$ = 1 for all $K^*$. Below the knee point location, all the stresses are seen to increase with streamwise distance in the APG region, while above the knee point, the stresses are seen to decrease. In the u-RS shown in figure \ref{fig:steady_stats}b, this increase and decrease in magnitude below and above the knee point, respectively, cause the knee point to become indistinguishable for $x/L_x \geq$ 0.8. In the v-RS and uv-RS shown in figures \ref{fig:steady_stats}(c,d), the increase in magnitude below the knee point with increasing $x/L_x$ yields a dramatic increase in the magnitude of the first peak (by factors of 2.8 and 4.4, respectively, for $K^*$ = 1). For $x/L_x >$  0, the Reynolds stresses in figures \ref{fig:steady_stats}b-d are weaker for larger $x/L_x$ or higher $K^*$ over most of the boundary layer, despite larger $x/L_x$ indicating longer exposure to an APG and higher $K^*$ corresponding to a stronger APG imposition. For example, at $x/L_x$ = 1, the u-RS magnitude at $y = 0.3 \delta$ is 43$\%$ lower at $K^*$ = 1 compared to at $K^*$ = 0.

Overall, there are two key trends observed in the Reynolds stresses. One, the Reynolds stresses of the current APG TBL exhibit knee points and multiple peaks. Two, the trends with which the magnitude of Reynolds stresses vary with increasing APG strength along $x$ and $K^*$ are different from (in fact, opposite to) well-established APG effects. The appearance of knee points and multiple peaks in the second order statistics seen here have previously been noted in studies of TBLs that encounter sudden changes in boundary or freestream conditions. An internal layer within the TBL is said to have formed in such situations, resulting in similar inflectional statistical footprints. A discontinuity in surface curvature \citep{matai2018flow}, a step change in surface roughness \citep{hanson2016development}, a change in the sign of pressure gradient encountered \citep{tsuji1976turbulent}, the application of blowing/suction \citep{simpson1971effect}, can all trigger the formation of internal layers, which have been shown to dominate the near-wall dynamics of the TBL, especially through skin friction. In the present investigation, the only change in freestream conditions occurs at $x'/L$ = 0.5 or $x/L_x = 0$, where the sense of the pressure gradient changes from favorable to adverse. An internal layer is inferred to have formed there from the change in sign of the pressure gradient, giving rise to the inflectional Reynolds stress profiles observed. \citet{uzun2021simulation} and \citet{baskaran1987turbulent} have similarly observed an internal layer triggered within the TBL at the favorable-adverse pressure gradient switch at the apex of the bump / hill geometry, leading to inflectional stress profiles at and downstream of the apex. 

\begin{figure}
    \centering
    \captionsetup{width=1\linewidth}
    \includegraphics{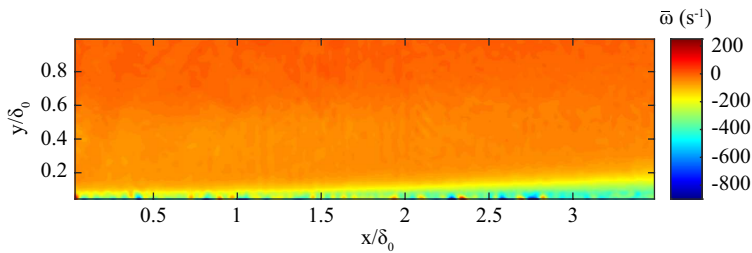}
    \caption{Time-averaged vorticity field for $K^*=1$. The average vortex activity is seen to be concentrated within a region close to the wall (presumably within the internal layer), outside which a slow suppression is seen along the streamwise direction.}
    \label{fig:vort}
\end{figure}

The formation and growth of the internal layer in the present investigation can be visualized using the average vorticity in the flowfield. Figure \ref{fig:vort} shows the time-averaged spanwise vorticity field for $K^*$ = 1. The spatial dimensions have been normalized by the ZPG boundary layer thickness. A concentrated region of strong vorticity is observed close to the wall. This region / layer is seen to grow in the APG along with the boundary layer thickness, but at a different growth rate. The average spanwise vorticity at $K^*$ = 0 (ZPG) does not show a distinct layer similar to the one here. The demarcation between the two layers in figure \ref{fig:vort} approximately matches with the local wall-normal location of the knee point in the stress profiles (figures \ref{fig:steady_stats}b-d), allowing the knee point to be considered the approximate edge of the internal layer. This definition is similar to those used by \citet{tsuji1976turbulent} and \citet{baskaran1987turbulent} for their internal layers. Within this edge, figure \ref{fig:vort} shows the vorticity magnitude increasing with $x$, and outside the layer and within the TBL, shows it decreasing with $x$. 

It is conjectured that the presence of the internal layer, along with the upstream FPG, is related to the counter-intuitive Reynolds stress variations observed with increasing APG strength along $x$ and $K^*$. Along $x$, the outer-scaled stresses increase near the wall and decrease away from the wall, and with increasing $K^*$, they decrease throughout most of the boundary layer.  The trends along $x$ are consistent with the data depicted in the literature of TBLs in APG regions of bump/hill flows that have undergone a preceding FPG region, showing a strengthening inner peak and a weakening/frozen outer peak under the APG \citep{baskaran1987turbulent,balin2021direct,cavar2011investigation,matai2019large}. Referring back to figure \ref{fig:vort}, the strengthening first peaks in figures \ref{fig:steady_stats}b-d (resolved in the v-RS and uv-RS, but inferable from u-RS) could be directly attributed to the formation and growth of a new layer with significant turbulent activity. The reason for the suppression of stresses outside this layer, however, is not obvious. Up until the exit of the FPG region (or until $x/L_x$ = 0), the TBL has experienced a strongly stabilizing effect that is known to be destructive to the large scale motions in the outer region \citep{harun2011structure}. This is indeed manifested by the suppressed stresses at $x/L_x$ = 0 in figures \ref{fig:steady_stats}b-d. The succeeding APG in $x/L_x > 0$, however, might have been expected to encourage the recovery of the turbulent stresses in the outer region through energized large scale motions. Instead, a further suppression of stresses is observed with $x$. When an internal layer is formed within a TBL, the boundary layer is said to be bifurcated, with the inner and outer layers developing independently of each other \citep{baskaran1987turbulent}. It is possible that the current internal layer deprives the outer flow of key energy transfer mechanisms that energize large scale motions. In their analysis of a TBL encountering a step change in surface roughness, \citet{hanson2016development} saw a similar reduction of energy contained in the outer region in their statistical and spectral results and noted that the internal boundary layer caused significant changes throughout the flow, including a ``shielding" of the outer region by the internal layer. A similar mechanism could be in play in the current experiments, to result in the reduction of stresses outside the internal layer along the streamwise APG. The observation of decreasing Reynolds stresses throughout most of the boundary layer with increasing APG strength with $K^*$ is likely a direct result of the stronger upstream FPG for cases with higher $K^*$. A stronger upstream FPG exerts a stronger downstream influence on the boundary layer through a stronger suppression of stresses, consistent with history effects being cumulative in nature. This upstream FPG effect lasts throughout the APG region recorded, with which the internal layer effect competes within the layer and aids outside the layer. This is why, at a given location in $x/L_x <$ 0.8, the stresses decrease with $K^*$ everywhere, but in $x/L_x \geq$ 0.8, the stresses strictly decrease with $K^*$ only in $y > ~ 0.1\delta$, as the internal layer competes to increase the magnitude of the first peak and aids in the reduction of stresses outside this peak. Further investigation into the role of upstream pressure gradients and internal layers in modifying the outer region behavior of FAPG (favorable-adverse pressure gradient) TBLs is warranted, especially given the consequential nature of outer region dynamics in APG TBLs \citep{lee2017large,monty2011parametric}.

\section{Conclusion}

An experimental campaign to study turbulent boundary layers (TBLs) experiencing families of pressure gradients characterized by a spatially-varying favorable and adverse pressure gradient (FAPG) in sequence was carried out. Planar particle image velocimetry was performed in the APG region of the flow for 22 pressure gradient conditions. Mean and Reynolds stress statistics are presented here for 6 of the 22 cases, while the data of all cases are provided as supplemental material, along with the pressure distributions and boundary layer edge parameters. The upstream favorable pressure gradient (FPG) had expected influences on the mean profiles within the adverse pressure gradient (APG): The mean velocity field was accelerated by the upstream FPG and then somewhat decelerated by the local APG. However, the Reynolds stresses showed results that, while consistent with prior work on bump flows, were qualitatively different from traditional expectations of APG TBLs. The stresses showed a bimodal structure coming out of the spatially-varying FPG and showed significant growth of the inner region and decay of the outer region as the flow advanced through the APG region, particularly for stronger pressure gradients. These findings suggested the formation of an internal layer that qualitatively changed the response of the boundary layer due to the change in sign of the pressure gradient applied. It was found to be more appropriate to study the resulting statistical evolution in terms of regions inside and outside this layer, rather than in terms of the traditional inner and outer regions of a TBL, especially since the decisive internal layer occupied a significant portion of what would usually be defined as the outer region. The qualitative agreement between the statistics of the present work and that of bump / hill flows, despite the absence of surface curvature in the present work, raises interesting questions about the interplay of pressure gradient and curvature effects on TBLs. Future work is encouraged to better understand the formation of the inner layer and the implications of the work for computational prediction and the physics of separation under an APG with an upstream FPG. 

\backsection[Supplementary data]{\label{SupMat}The statistics for all cases are included as a MATLAB\textregistered  \hspace{0.5mm} file in HDF5 format.} 
\backsection[Funding]{This work was supported by the Office of Naval Research through grant \# N00014-21-1-2648. Support from the Grainger College of Engineering and the Aerospace Engineering Department at the University of Illinois Urbana-Champaign is gratefully acknowledged.}
\backsection[Declaration of interests]{The authors report no conflict of interest.}

\bibliographystyle{jfm}
\bibliography{jfm}

\end{document}